# Configurational Entropy of Folded Proteins and its Importance for Intrinsically Disordered Proteins

Meili Liu[1-3], Akshaya K. Das[2,3], James Lincoff[3,4], Sukanya Sasmal[3,4], Sara Y. Cheng[2,3], Robert Vernon[6], Julie Forman-Kay[6,7] and Teresa Head-Gordon[2,3,4,5][†]

[1]Department of Chemistry, Beijing Normal University, Beijing 100875, P. R. China

[2]Department of Chemistry and [3]Pitzer Center for Theoretical Chemistry, [4]Department of Chemical and Biomolecular Engineering, [5]Department of Bioengineering, University of California, Berkeley 94720 USA

[6]Molecular Structure and Function Program, Hospital for Sick Children, Toronto, Ontario M5G 0A4, Canada
[7]Department of Biochem., University of Toronto, Toronto, Ontario M5S 1A8, Canada

Many pairwise additive force fields are in active use for intrinsically disordered proteins (IDPs) and regions (IDRs), some of which modify energetic terms to improve description of IDPs/IDRs, but are largely in disagreement with solution experiments for the disordered states. We have evaluated representative pairwise and many-body protein and water force fields against experimental data on representative IDPs and IDRs, a peptide that undergoes a disorder-to-order transition, and for seven globular proteins ranging in size from 130-266 amino acids. We find that force fields with the largest statistical fluctuations consistent with the radius of gyration and universal Lindemann values for folded states simultaneously better describe IDPs and IDRs and disorder to order transitions. Hence the crux of what a force field should exhibit to well describe IDRs/IDPs is not just the balance between protein and water energetics, but the balance between energetic effects and configurational entropy of folded states of globular proteins.

[†]corresponding author: thg@berkeley.edu

# INTRODUCTION

Intrinsically disordered peptides (IDP) are a class of proteins that are defined as dynamic structural ensembles rather than a dominant equilibrium structure in solution.[1] Experimental methods such as nuclear magnetic resonance (NMR) spectroscopy[2], single molecule fluorescence Förster resonance energy transfer (smFRET)[3], and small angle x-ray scattering (SAXS)[4] can provide restraints on the structural ensemble of IDP systems, but are unable to fully resolve important sub-populations of structure relevant for function.[5] Therefore computational methods play a critical role by first generating putative structural ensembles[6] and secondly reconciling them with the highly averaged experimental information using Monte Carlo optimization[7,8] or more recently Bayesian formalisms[9,10,11]. In this work we are concerned with the generation of IDP ensembles using physically motivated force fields and molecular dynamics simulations (MD) which model protein-protein, protein-water, and water-water interactions at the atomic level.

Nearly all MD simulations of IDP structural ensembles have been generated with pair-wise additive force fields which have traditionally been parameterized to reproduce the folded states of proteins.[12] Nonetheless, atomistic force fields have struggled with issues ranging from biases in secondary structure conformations[13,14] or overly-structured and collapsed ensembles that do not agree with experimental data on many IDP systems.[15,16] Additionally, IDPs are more solvent-exposed than folded globular proteins, thus the corresponding choice of water model used to simulate IDPs is critical for capturing the correct balance between protein-water and water-water interactions for folded and unfolded states and for disordered proteins.[2,17,18] The D.E. Shaw group were also the first to show that long standard MD simulations - on the order of hundreds of microseconds - are required to ascertain the ability of a force field to maintain the structural integrity of a globular protein.[19,20] We found that similar issues arise for IDPs that also require long simulations and/or accelerated sampling methods to better represent their structural properties.[21]

To improve upon MD simulated predictions for IDPs, a few research groups have proposed parameter changes to standard force fields to bring them better in line with solution experiments. For the TIP4P-D water model[22], Piana et al. increased the $C_6$ dispersion coefficient of the Lennard-Jones parameter by ~50% to make London dispersion interactions more favorable, and when combined with Amberff99sb-ildn model[19] for the protein, resulted in more expanded IDPs with improved agreement with experimental NMR and small angle X-ray scattering (SAXS) data. Best and Mittal[23] introduced backbone parameter modifications of one of the Amber force fields combined with the TIP4P/2005 water model[24] to reproduce, for example, the temperature dependence of the helix-coil transition for the 15-residue peptide Ac-(AAQAA)$_3$-NH$_2$ peptide. The resulting A03WS/TIP4P/2005 is intended for use for IDPs, but when applied to poly-glutamine IDP in solution was found to generate mostly featureless and highly extended conformations that do not correctly describe solution experiments.[25] Independently, Henriques et al. have shown that both Amberff99sb-ildn/TIP4P-D and A03WS/TIP4P/2005 reproduce better radius of gyration values for the disordered Histatin 5 (Hst 5) peptide, although both force fields exhibit more turn content for Hst 5 that creates slightly more collapsed states.[15] Robustelli et. al. performed extensive millisecond MD simulations on six different pairwise additive protein force fields on a range of fully-disordered to folded globular protein systems.[20] These simulations revealed that none of these standard force fields agreed with experimental data for a number of IDP systems while also maintaining the ability to accurately model folded proteins.[20]

Therefore newer protein force fields and water model combinations have been proposed to capture the behavior of IDPs as well as folded proteins.[12] This is important for at least two reasons. First they can be used when simulating interactions of IDPs with folded proteins[26], disorder-to-order transitions[27], and folded proteins with intrinsically disordered regions (IDRs)[28]; second they satisfy the goal of any force field, which is transferability to new protein systems and other emerging problems such as liquid phase separation[29]. An example is the CHARMM36m protein model of Huang et. al. that purports to better describe both IDPs and folded proteins using the same set of refined peptide backbone parameters and salt-bridge interactions, and an increased Lennard-Jones (LJ) well depth to strengthen protein-water dispersion interactions.[30] These modifications led to a reduction in the percentage of predicted left-handed α–helices as well as better agreement with NMR scalar couplings and SAXS curves for folded proteins, although Huang et. al. observed that no universal interaction strength parameter in the Lennard-Jones function could generate structural ensembles with good agreement with experimental radius of gyration measurements for all IDP systems[30]. Hence the logical next step is to consider more advanced potentials that can be made more accurate by including multipolar electrostatic interactions with many-body polarization that can respond to changes in the solvent conditions around biomolecules.[31, 32]

One purpose of this study is to ascertain how well the many-body polarizable AMOEBA protein (AmPro13)[33] and water (AmW03)[34] force field performs against experiments across of range of folded proteins, IDRs and IDPs, when compared to a representative standard force field, AMBERff99sb/TIP3P(TIP4p-Ew), and recently modified fixed charge force fields, CHARMM36(m)/TIP3P(m), where the parentheses refer to alternate protein and/or water model combinations. A second important purpose of this work is to provide some easily ascertained measures of what constitutes a successful force field that can simultaneously describe both folded proteins and proteins with disorder.

We first show that all force field combinations yield stable trajectories over the 1 μs simulation time for the 7 globular proteins ranging in size from 130-266 residues. But an important distinction is that the polarizable force field exhibits substantially larger root mean square deviations (RMSDs) and fluctuations (RMSFs) than that of the non-polarizable models, although all force fields maintain an average radius of gyration $<R_g>$ in agreement with experiment and structural similarity criteria[35] consistent with a correctly folded structure. By examining the RMSFs using the Lindemann criteria[36, 37], we show that the fixed charge force fields simulate folded protein models that are more solid-like throughout their structure, whereas the polarizable model displays greater fluidity with Lindemann values close to recent inelastic neutron experiments.[37]

We hypothesized that a force field that provides the largest structural deviations and statistical fluctuations that remains consistent with the experimental $R_g$ of a folded globular protein, will better be able to capture the greater plasticity and match solution experiments for IDPs and IDRs. In fact we consistently find that the polarizable model better reproduces the experimental $R_g$[38] for the disordered Hst 5 peptide, exhibits a stronger temperature dependence in the disorder to order transition for the (AAQAA)$_3$ system due to unusual π–helical structure, and maintains a folded core for the TSR4 domain while simultaneously exhibiting regions of disorder. By contrast the fixed charge force fields have $R_g$ distributions that are in disagreement with SAXS intensity profiles and contain higher populations of turns for Hst 5 that contribute to a more collapsed state, and they show little change with temperature for (AAQAA)$_3$. Our results emphasize the importance of configurational entropy for folded states, and determine a range of metrics for its evaluation, as a key factor for whether a force field will exhibit better predictive capacity for IDPs/IDRs.

## RESULTS

Figures 1 shows the 7 folded proteins we have considered in this study, in addition to the TSR4 domain (1VEX) as an intermediate class of protein with a small folded core dominated by IDRs.

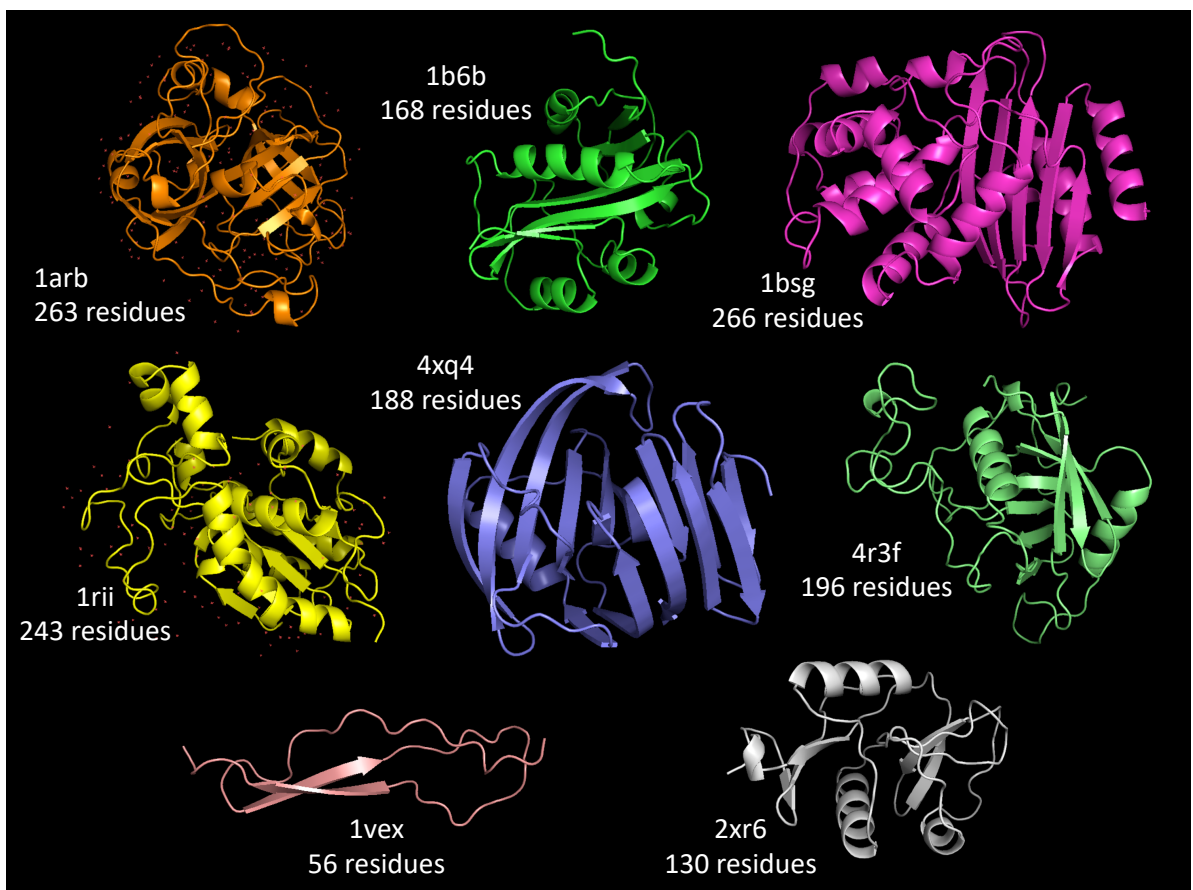

**Figure 1.** *Seven folded proteins and TSR4 (1VEX) simulated with polarizable and non-polarizable force fields.* (PDB IDs: 1B6B[39], 1ARB,[40] 1BSG[41], 1RII[42], 1VEX[43], 2XR6[44], 4R3F[45], and 4XQ4[46]).

Figures 2 and S1 report on the coordinate RMSD of the 7 folded proteins over the 1 μs of MD simulation for each of the force field combinations. All 7 folded globular proteins show no evidence of early unfolding events or significant degradation in secondary structure, although there is some upward drift for a handful of proteins for all force fields that we attribute to a limitation of the 1 μs simulations. Although our 1 μs simulation timescales are typical of previous work on measuring protein stability[30], we consider additional metrics for acceptable deviations from the starting structures derived from the PDBs.

Figure 2 reports a metric developed by Maiorov and Crippen that provides an empirical relationship to estimate structural similarity $D_{0,sim}$ and dissimilarity $D_{0,dis}$ for globular proteins (see Table S1 and S2).[35] Values below or at the $D_{0,sim}$ similarity measure defines a valid ensemble of structures for which loop regions may reconfigure while not significantly shifting the $<R_g>$ and core fold, while values at or above the $D_{0,dis}$ metric distinguishes the dissimilarity between a reference structure and its mirror image and thus any large shifts in $<R_g>$ and conformation. In this work we measure $R_g$ from both the PDB structure for each protein as well as from polymer scaling law estimates parameterized by PDB structures (see Table S2) under poor solvent conditions and structural variations of globular proteins of the same size.[47, 48] The larger $R_g$ values from the polymer scaling laws relative to the PDB structure are well within the expectations from solution experiments[49], and consistent with crystal structures differing somewhat from NMR[50] and SAXS[51] ensembles for folded states (Tables S1 and S2).

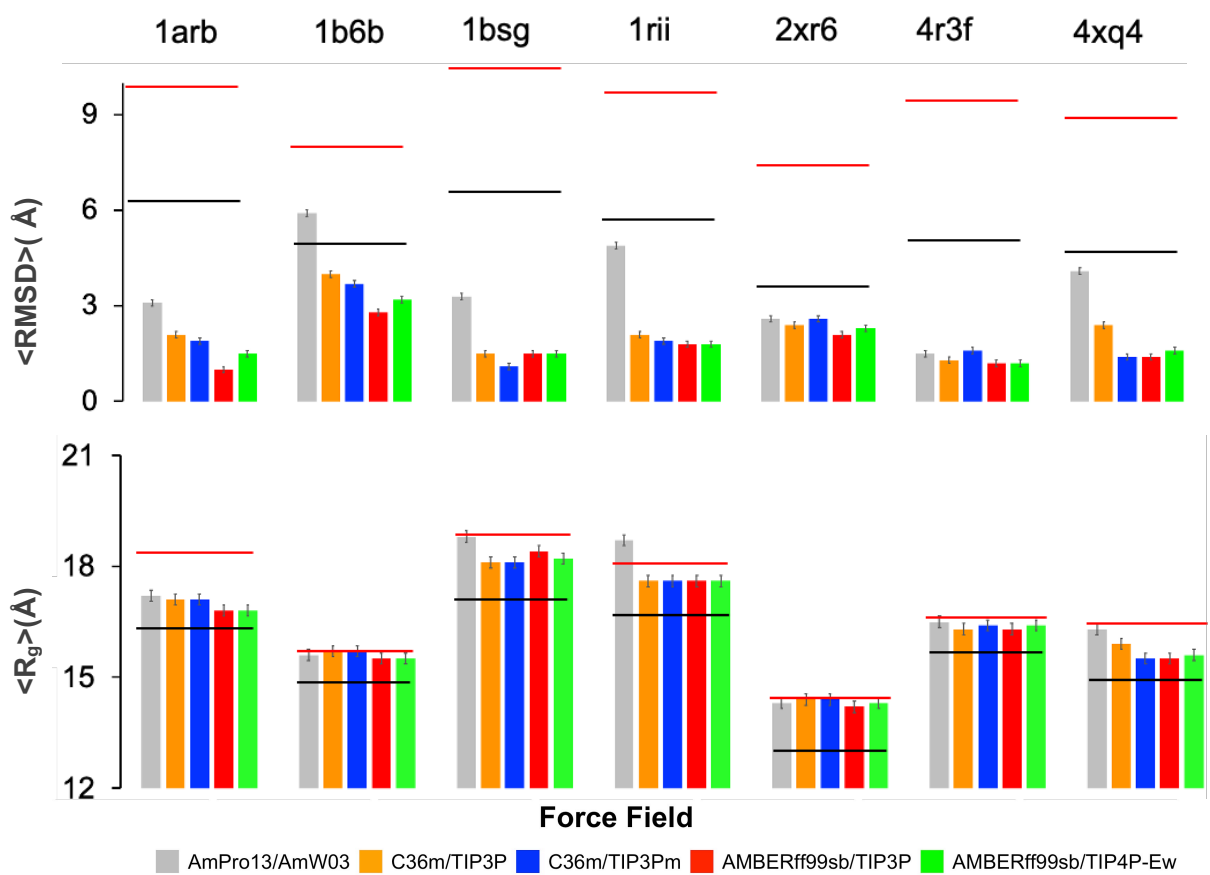

**Figure 2.** *Measures of protein stability when simulated with polarizable and non-polarizable force fields.* (a) Root mean square deviation (RMSD) for 1 μs MD simulations for AmPro13/AmW03, C36m/TIP3P, C36m/TIP3Pm, ff99SB/TIP3P, and ff99SB/TIP4P-Ew. The black line is the value of the $D_{0,sim}$ metric and red line the $D_{0,dis}$ metric. (b) $<R_g>$ for all force fields and comparison to the $R_g$ of the PDB structure (black) or polymer scaling laws (Table S2) as a measure of solution (red). Proteins characterized are 1ARB,[40] 1B6B[39], 1BSG[41], 1RII,[42] 4XQ4[46], 4R3F,[45] and 2XR6[44].

As seen in Figure 2, all force fields yield RMSDs within the range of $D_{0,sim}$ metric for the 7 folded proteins. With the exception for 1b6b for which the <RMSD> using AmPro13/AmW03 is within the $D_{0,sim}$ by ~0.5 Å, all models have not fully reached allowed values of the $D_{0,sim}$ metric, and no force field exhibits unfolding or instability as measured by $D_{0,dis}$ (see Figure S2). But just as importantly, it is also evident that the fixed charge force fields generally yield folded states with much smaller <RMSD> values, whereas the polarizable force field model is closest to the upper bound of the similarity metric for the globular proteins. In addition, the $<R_g>$ for the pairwise additive models are more often closer to the PDB structure, while the $<R_g>$ values for the polarizable model are more in line with polymer scaling law estimates (Table S1).

Because values of RMSD correlate directly with root mean square fluctuations (RMSF)[52], we compare the force fields using the Lindemann criterion developed originally for the melting of a solid Cryst.[53] The Lindemann value $\Delta_L = RMSF/a$ has been adapted to the case of proteins by replacing the crystal lattice constant, $a$, with an average non-bonded distance.[36, 37] Katava et al. provided experimental estimates of the RMSF from inelastic neutron scattering for chicken egg white lysozyme (CEWL), and assuming $a$ = 4.75 Å, found a Lindemann value at the protein melting temperature ($T_m$) of $\Delta_L^{exp}(T_m)$~0.17-0.18, driven by the mixing in of a greater proportion of unfolded state fluctuations.[37] Below $T_m$ the contributions from unfolded state fluctuations diminish as temperature decreases, of course, but Zhou et al showed that the folded state fluctuations are comprised of an interior protein core that is suppressed and solid-like ($\Delta_L^{core}$ ~0.05-0.1) whereas the protein surface is quite fluid ($\Delta_L^{core}$ ~0.15-0.2),[36, 54] which in part explains the overall experimental value for the CEWL protein near 300 K of $\Delta_L^{exp}(300\ K)$~0.15-16 in water solvent[37]. Because Katava and co-workers found similar results for myoglobin, crambin, hemoglobin, and BSA, they expect these results to be universal values for any folded state of a globular protein of average size, and hence we rely on comparisons to $\Delta_L^{exp}(300\ K)$ in our simulations of the 7 folded proteins analyzed here.

Table 2 reports the corresponding $\Delta_L^{sim}(300\ K)$ values for each protein, assuming a value $a$ = 4.375 Å which is an average taken among all previous work,[36, 37, 54] but with the RMSF calculated from the fixed charge and many-body force fields simulations (Table S3). Averaged over all of the folded proteins, the non-polarizable force fields yield Lindemann values $\Delta_L^{sim}(300\ K)$ of ~ 0.12; to put this value into perspective for the fixed charge force fields, this value is close to $\sim\Delta_L^{exp}(230\ K)$ for CEWL. By contrast, the polarizable force field predicts <RMSF> values that are ~30% larger than those of the fixed charge models, with values of $\Delta_L^{sim}(300\ K)\sim0.16$ that are in good agreement with the experimental value at room temperature. Table S4 shows that all force fields have a very solid structural core, $\Delta_L^{core}(300\ K)\sim0.09$ for the fixed charge force fields and ~0.12 for the polarizable model, and that their total simulated averages are thus dominated by their surface fluctuations, $\Delta_L^{surf}(300\ K)$, which are largest for the many-body potential (0.155 vs 0.21). The lower $\Delta_L^{sim}(300\ K)$ values from the fixed charge force fields are thus indicators that they will generally overestimate the melting temperature, an undesirable feature of standard force fields noted previously by Lindorff-Larsen and co-workers[55], because they do not fully activated their large-scale collective modes permitted by $D_{0,sim}$.

**Table 2:** *Lindemann values for 7 folded proteins at 300 K.* A value of $a$ =4.375Å and <RMSF> averaged over all residues (Table S2) were used to calculate the Lindemann value.

| Force Field/Protein | $\Delta_L^{sim}(300\ K)$ | | | | | | | |
|---|---|---|---|---|---|---|---|---|
| | 1bsg | 1arb | 1rii | 4r3f | 4xq4 | 1b6b | 2xr6 | Average |
| ff99sb/TIP3P | 0.14 | 0.10 | 0.13 | 0.11 | 0.12 | 0.14 | 0.11 | 0.12 |
| ff99sb/TIP4P-Ew | 0.12 | 0.10 | 0.12 | 0.12 | 0.10 | 0.13 | 0.11 | 0.11 |
| C36m/TIP3P | 0.11 | 0.11 | 0.12 | 0.12 | 0.14 | 0.14 | 0.11 | 0.12 |
| C36m/TIP3Pm | 0.11 | 0.12 | 0.13 | 0.12 | 0.12 | 0.18 | 0.14 | 0.13 |
| AmPro13/AmW03 | 0.18 | 0.13 | 0.22 | 0.16 | 0.17 | 0.16 | 0.13 | 0.16 |

We therefore anticipate that $T_m$ values using the polarizable force field will be in better agreement with experiment, since large surface fluctuations are evident by their $D_{0,sim}$ values that approach the estimated upper bound[35], while remaining consistent with the folded $R_g$. We thus conclude from the folded protein class that force fields should exhibit, in addition to a balance between protein-protein and protein-water energetics, a good balance between energy and configurational entropy in order to realize $\Delta_L^{sim}\sim\Delta_L^{exp}$.

We carry this idea further to predict that the force fields with $\Delta_L^{sim}\sim\Delta_L^{exp}$ for folded proteins will be better suited to representing the structural ensembles of IDRs and IDPs as well; by corollary, force fields with $\Delta_L^{sim}<\Delta_L^{exp}$ for folded states will not be able to describe the greater plasticity of intrinsically disordered states. To test the extrapolation from folded proteins, we now consider the TSR4 domain, which is comprised of a small β-sheet core stabilized by a network of pi-contacts, with large loops that have been classified as intrinsically disordered regions.[56] For TSR4 the <RMSD> for all force fields (Table 3) are well outside the $D_{0,sim}$ metric (1.34 Å), and in better agreement with the $D_{0,dis}$ value (4.49 Å) given the presence of significant segments of disorder. For the TSR4 domain all force fields have a less solid structural core than for the folded proteins, $\Delta_L^{core}$~0.16-0.18 and are dominated by large surface fluctuations $\Delta_L^{core}$~0.18-0.29 that exceed that of the folded proteins. There is no direct solution experimental data to validate against, but these results support the expectation that the Lindemann criteria value for globular proteins is not universal and cannot be extended to IDRs and IDPs.

**Table 3.** *Fluctuation properties of the TSR4 domain at 300 K.* <RMSD> is the average root mean square distance to starting structure of TSR4. A value of $a$ =4.375Å and <RMSF> averaged over all residues of TSR4 were used to calculate the total Lindemann value, $\Delta_L^{sim}$. $\Delta_L^{core}$ was evaluated from the β–sheet core residues; $\Delta_L^{surf}$ was calculated from all protein residues not characterized as core residues.

| Force Field | $<RMSD>$ | $\Delta_L^{core}$ | $\Delta_L^{surf}$ | $\Delta_L^{sim}$ |
|---|---|---|---|---|
| ff99sb/TIP3P | 3.8 | 0.16 | 0.18 | 0.17 |
| ff99sb/TIP4P-Ew | 3.5 | 0.16 | 0.20 | 0.18 |
| C36m/TIP3P | 3.1 | 0.17 | 0.23 | 0.20 |
| C36m/TIP3Pm | 3.0 | 0.18 | 0.24 | 0.21 |
| AmPro13/AmW03 | 5.5 | 0.20 | 0.29 | 0.25 |

We also find that the Amber force fields yield the most suppressed $\Delta_L^{sim}(300\,K)$ values, while the C36 and C36m force fields fluctuate more, and the polarizable model yields the largest $\Delta_L^{sim}(300\,K)$ value for the TSR4 domain. These significant $\Delta_L^{sim}(300\,K)$ differences for the TSR4 domain would lead to substantial differences among the force fields with complete disorder. We therefore next consider Histatin 5, a cationic IDP for which it has been challenging using fixed-charge force fields to achieve agreement with the reported experimental data. These include SAXS form factors that measure a $<R_g>$ = 13.8 ± 2.2 Å[38], and solution CD and NMR[57, 58] measurements showing that Hst 5 lacks significant secondary structure in aqueous solution, although Hst 5 prefers α-helical conformations in non-aqueous solvents. From Figure 3 we see that the pairwise additive force fields ff99SB/TIP3P, C36m/TIP3Pm, and C36m/TIP3P predict a more narrow $R_g$ distribution around compact structures with $<R_g>$~10.0-11.0 Å, with higher populations of turns that likely account in part for these collapsed states; the ff99SB/TIP4P-Ew model predicts a bimodal distribution of collapsed and expanded states, but this is in disagreement with the SAXS form factor. The AmPro13/AmW03 potential, with no force field modifications, predicts a more expanded $<R_g> \sim 14.0$-$14.5$ Å in good agreement with the SAXS observable and NMR and CD experiments.

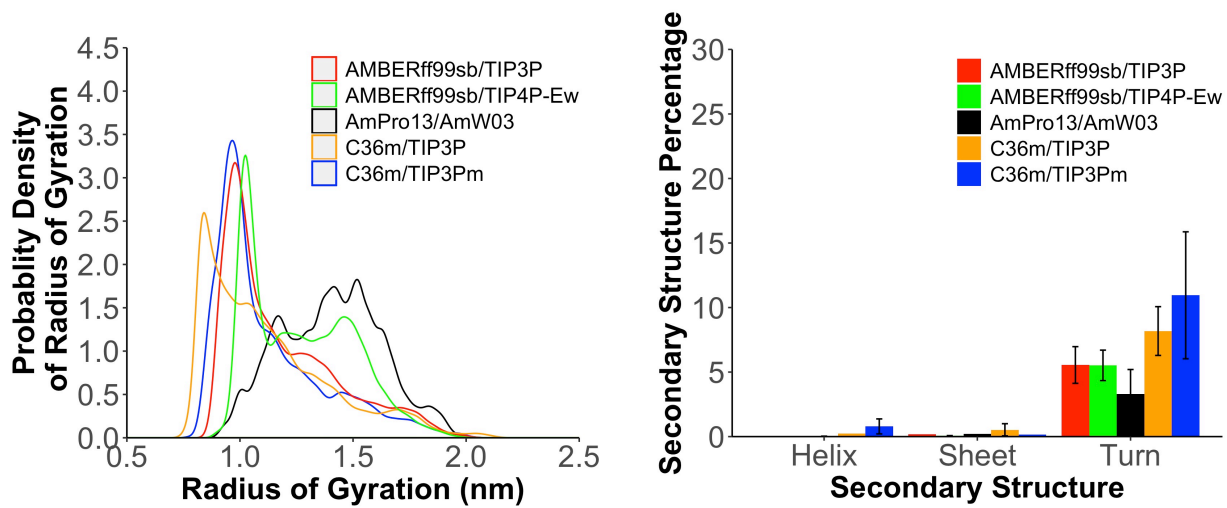

**Figure 3:** *Structural properties for Hst 5 using polarizable and non-polarizable force fields.* (a) Probability density estimates of the radius of gyration and (b) average percentages of different secondary structures features for the disordered Hst 5 peptide.

Finally, we consider the very challenging temperature dependence of the (AAQAA)$_3$ peptide, in which NMR experiments have previously ascertained a (partial) disorder to order transition as temperature is lowered. There are several issues that are not sufficiently discussed in the literature regarding this peptide and previous simulation attempts to reproduce its behavior. First is that the NMR experiment was designed to determine the $^{13}$C-carbonyl shift at each residue, providing an experimental measure of the helicity at each residue for comparison to a helix-coil model that predicts the helicity at each residue.[59] Hence an overall percentage averaged across all 15 residues is not the correct measure as the NMR shifts are residue-specific values, yielding estimates of 0% to 25% depending on position, with the N-terminus being more helical; this is in contrast to the highly symmetric prediction of the helix-coil model.[59] Finally, previous studies that found that alanine peptides are unusually enriched[60, 61] with π-helix in particular, while the $^{13}$C-carbonyl chemical shifts are not generally able to differentiate among all three helix categories, especially for fluctuating states. Note that there are statistically different shifts for stable α–helix and 3$_{10}$ helix[62], and suggests that comparison of structural ensembles to the standard NMR experiment should combine the propensities of the different helix types.

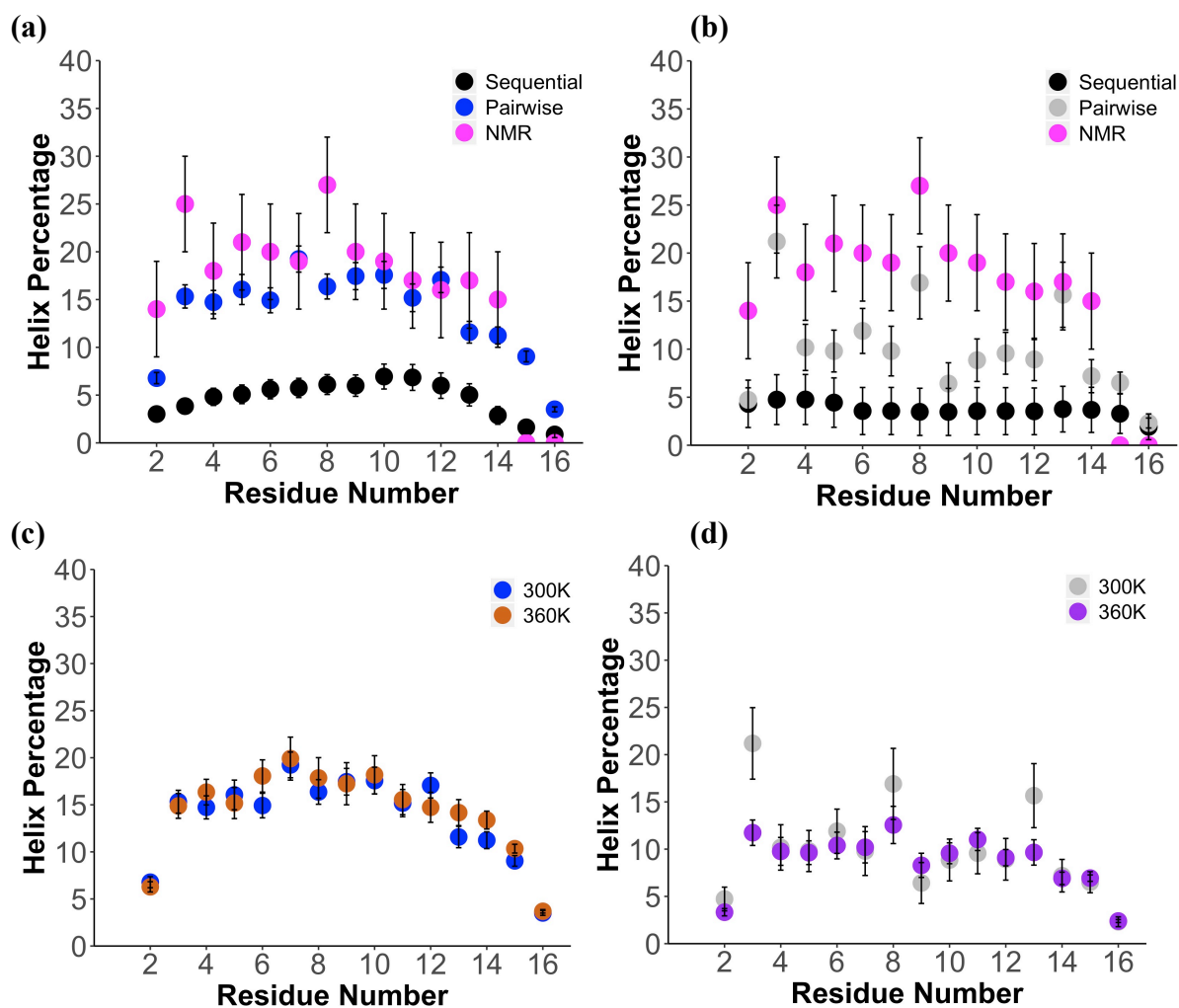

**Figure 4:** *Structural properties for (AAQAA)$_3$ using polarizable and non-polarizable force fields.* Comparison of estimated helical propensities from NMR (pink), average α–helix from simulation assuming 3 sequential residues (black), and pairwise average over any presence of α–helix, π–helix, and 3$_{10}$ helix for (a) C36m/TIP3Pm (blue) and (b) AmPro13/AmW03 (gray) at 300 K. Comparison of changes in helix propensity with temperature at 300 K and 360 K for (c) C36m/TIP3Pm and (d) AmPro13/AmW03.

We first investigate the definition of an α–helix percentage used by previous research groups, defined as 3 consecutive residues residing in a broad α–helix basin of the Ramachandran plot (labeled sequential in Figure 4). Unlike more recent studies, we provide individual residue percentages for the (AAQAA)$_3$ peptide (Figure 4a,b and Figure S3).[63] As determined by Boostra and co-workers[64], the C36m results depend critically on the "right" water model, i.e. the standard TIP3P water model must be used, to predict the higher helical content at low temperatures, with little helical content observed using TIP3Pm at any temperature. We support that result using TCW sampling in which C36m/TIP3Pm yields ~5% α–helix at 300 K (Figure 4a) as do the other fixed charge force fields (Figure S3), and they all exhibit a flat temperature dependence (Table S5), in very good agreement with Robustelli et al using 20 μs MD simulations.[20] The AmPro13/AmW03 polarizable model gives α–helical percentages that are similar to the Amber and CHARMM force fields for (AAQAA)$_3$ peptide, i.e. < ~5% with no disorder to order transition (Figure 4b and Table S5).

Instead we consider an alternative definition of helical percentages in which the (AAQAA)$_3$ peptide might adopt not only α–helix, but π–helix and $3_{10}$ helix configurations[60] as well based on values of ψ(i) and φ(i+1) values (which we label pairwise in Figure 4). Figures 4a, 4b and S3 show that, when using this definition, the fraction of helical percentages for each residue increases for all force fields and temperatures, ~ 15-20%, but with important differences between the polarizable and non-polarizable models. It is seen that the fixed charge models (Figures 4c and S3) have no temperature dependence, with nearly the same helical percentages at 300 K and 360 K. By contrast, the AmPro13/AmW03 model shows some temperature dependence, with a loss of helical structure at 360 K relative to 300 K as seen in Figure 4d. This supports our hypothesis that fixed charge force fields that are overly stabilized for folded proteins will manifest as too inflexible for disordered states, in this case due to the inability to melt the N-terminal helix of (AAQAA)$_3$ at high temperatures, unlike the polarizable model which exhibits a better temperature dependence for the configurational entropy. This result has also addressed a long standing problem with the characterization of the (AAQAA)$_3$ peptide with temperature using simulation, that must emphasize not only standard α–helix, but π–helix and $3_{10}$ helix categories as well, and to characterize not average helix percentages over the whole peptide, but the residue-by-residue average helical percentage values instead.

**DISCUSSION AND CONCLUSION**

We have presented a comparison of a range of pairwise additive force fields and the many-body force field AMOEBA to test their ability to simultaneously describe stable folded states of 7 globular proteins, proteins with regions of disorder illustrated with the TSR4 domain, the Hst 5 IDP, and the partial disorder to order transition as temperature is lowered for the (AAQAA)$_3$ peptide. We find that the fixed charge force fields yield small RMSD differences from the PDB structures of the folded globular proteins, whereas the polarizable model has larger RMSD values that are within the expectations from solution experiments[49, 50, 51] on folded states. However, we have also shown that force fields that generate the largest RMSDs that are still consistent with the experimental $R_g$, and thus exhibiting larger statistical fluctuations on average, are better able to simultaneously describe the plasticity of proteins with regions or complete structural disorder as shown for the TSR4 domain, Hst 5, and the (AAQAA)$_3$ peptide.

In particular, the polarizable AMOEBA force field presents a significant advance over fixed-charge force field for IDP simulations, even those that have been specifically modified to better reproduce IDP behavior, as it does not require any problem-specific parameterization for IDPs, and can be used as a general force field for different types of IDPs and their complexes. Our analysis indicate that the fixed charge force fields uniformly describe overly collapsed and rigid structural ensembles of the folded proteins, whereas the polarizable model is inherently more fluid with greater configurational entropy that captures both folded structure and structural ensembles of IDPs. Finally, we note that other force fields tested previously on (AAQAA)$_3$ should be reevaluated to consider both π–helices and 3$_{10}$ helices in addition to α–helix, with a metric that evaluates the helical content on a residue-by-residue basis as the C-terminal end remains unstructured at any temperature.[59] We also note that more current state-of-the-art estimates of helical structure based on NMR shifts could be used to obtain a better experimental reference for this peptide.[65, 66]

We believe that the analysis we have presented here offers several new ideas on force field validation criteria. The first is to measure the ability of a force field to more systematically approach the full value permitted by the structural similarity $D_{0,sim}$ metric for globular proteins[35], as well as a Lindemann criteria values $\Delta_L^{sim}$ that are close to that determined from inelastic neutron scattering experiments and that are touted to be a universal criteria for any folded protein in water[37]; a related metric is the ability to reproduce the melting temperature of folded proteins. These measures are best at assessing the balance between energetic effects and configurational entropy, and what a force field should exhibit to equally well describe IDRs/IDPs and folded states of globular proteins. While this study has concluded that the polarizable AMOEBA force field is better by these structural and dynamical metrics, it is still an open question as to whether some fixed charge force fields are capable to the same extent, or can be made more capable in this regard. While we found that the pairwise additive force field combinations examined here are not fully sufficient, further evaluation and fitting to reproduce the dynamical criteria introduced can provide good guidance to improving force fields in general.

# METHODS

The Hst 5, TSR4 domain, and the 7 folded protein systems were modeled with the following force field combinations: Amberff99sb (ff99SB)[67] with TIP3P[68] and TIP4P-Ew,[69] CHARMM36m (C36m)[30] with TIP3P[68] and Charmm modified TIP3P (TIP3Pm), and AmPro13[33] with Amoeba Water03 (AmW03)[34]. We used 1 microsecond standard MD simulations for the folded proteins, the TSR4 domain, and the Hst 5 system with the OpenMM[70] package for the fixed charge force fields and the Tinker-OpenMM platform[71] for AMOEBA. We also developed a modified version of the OpenMM[70] and Tinker-OpenMM platforms,[71] to perform calculations on graphics processing units (GPUs) with Temperature Cool Walking (TCW)[21, 72, 73] to further improve the sampling of the $(AAQAA)_3$ systems. For $(AAQAA)_3$, we considered the force field combinations of ff99sb/TIP4P-Ew, ff99sb-ildn/TIP4P-D, C36m/TIP3Pm, C36/TIP3Pm, and AmPro13/AmW03 models.

***System and Simulation Preparation***: Initial disordered state structures for Hst 5 and Ace-$(AAQAA)_3$-Nme were generated using the tleap function in the AMBER MD engine.[74] The initial coordinates of the TSR4 and seven folded proteins were taken from their PDB structures. Solvation of these systems were performed using tleap for simulations using the ff99sb force fields, VMD or the online CHARMM-GUI for simulations using the C36m force field,[75] and TINKER 8 for simulations using the AmPro13 force field.[76] All simulations were performed on systems with addition of $Na^+$ or $Cl^-$ counter-ions to maintain net zero charge.

The Hst 5 system was equilibrated according to the following procedure. First, the fully extended peptide was solvated using a 10 Å buffer, and the system was simulated at 500 K for 1 nanosecond (ns) in the NVT ensemble to collapse the peptide. Second, the peptide was re-solvated using a smaller cubic box with side lengths of 59.1Å, with a total 6166 water molecules. The re-solvated peptide was equilibrated with NVT conditions at 500K for 1 ns, followed by 1ns of NVT at 300K. Finally, the peptide was run in the NPT ensemble at 300K, to equilibrate the size of the simulation box. The initial structure for production NVT MD simulations was chosen based on the maximum probable density.

For the $(AAQAA)_3$ system the peptide was started from an α–helix and solvated using a 10 Å buffer, and the heavy atoms of the protein backbones were harmonically restrained with a spring constant of 10 kcal/mol/Å$^2$ during a 1 ns simulation in the NPT ensemble over a temperature range that captures the transition (300 K, 320 K, 340 K, 360 K, or 380K). Second, 100 ps of NPT simulations were run where the position restraints of the protein backbone were relaxed from 10.0 kcal/mol/Å$^2$ to 0.0 kcal/mol/Å$^2$, reducing the spring constant by 1.0 kcal/mol/Å$^2$ every 10 ps. Finally, 20 ps of NPT simulations were run with no restraints on the protein backbone.

Finally, the larger protein systems were energy minimized to a potential energy tolerance of 0.5 kJ/mol with a non-bonded cutoff of 9.4 Å. The heavy atoms in the protein backbones were harmonically restrained with a spring constant of 10 kcal/mol/Å$^2$, and the system was heated in the NVT ensemble from 10K to 300K at a rate of 1 K/ps using Langevin integrator with a 1 fs timestep. Once the systems reached a 300K, a 1 ns simulation was run in the NPT ensemble with a rRESPA multi-timestep integrator with a 4 fs timestep for fixed-charge force fields and 2 fs timestep for polarizable force fields, using an Andersen Thermostat at 300K with a collision frequency of 50 ps$^{-1}$. A Monte Carlo Barostat was used with a target pressure of 1.01325 bar with an exchange attempt frequency made every 50 fs.

***Production Simulation Details and Analysis***: For the solvated TSR4 and folded proteins, we performed 1 μs molecular dynamics simulations in the NVT ensemble at 300 K with the Bussi thermostat using the RESPA integrator and heavy-hydrogen mass repartitioning with a 3 fs time step. Ewald cutoffs of 7 Å and van-der Waals cutoff of 12 Å were used. A pairwise neighbor list for partial charge and polarizable multipole electrostatics and for van der Waals interactions was used. A grid size of 64Å x 64Å x 64Å was used for PME summation and a 10$^{-4}$ Debye convergence criterion for self-consistent induced dipoles. Frames were saved every 10 ps and used to perform further analysis. For (AAQAA)$_3$, the TCW simulations were performed in the NVT ensemble with the Andersen Thermostat and velocity verlet integrator with a 2 fs timestep to propagate the target temperature (300 K, 320 K, 340 K, 360 K, or 380 K) and high temperature (456 K) walkers. Frames from the low temperature replica were saved every 1 ps and used to perform further analysis.

Figure S1 shows the raw RMSD and RMSF over the 1 μs trajectory for the folded proteins. Analysis of the trajectories were performed using Amber Tools and in-house analysis scripts to analyze the secondary-structure propensity for Hst 5, radius of gyration for Hst 5 and the folded proteins, and/or RMSDs and RMSFs of the protein-water systems using block averaging over ~50ns blocks over the last 800 ns of the trajectory. For the (AAQAA)$_3$ system, a residue was classified as being in a helical conformation using two different definitions when compared with NMR chemical shift data from experiments.[59] The first definition is defined as a series of three consecutive residues where the φ dihedral angle was between −160° to −30° and the ψ angle was between −67º to −7º.⁶¹ The second definition more directly targeted different types of helices; when the first and last residue pairs are excluded, the ψ dihedral angle of one residue and the φ dihedral angle of the next residue sum to −125°±10° for the π-helix, −75°±10° for the $3_{10}$ helix, whereas that for the α-helix is −105°±10°.

**ACKNOWLEDGEMENT**. The Berkeley investigators thank the National Institutes of Health for support under Grant 5R01GM127627-03. M. Liu thanks the China Scholarship Council for a visiting scholar fellowship. J.D.F.-K. also acknowledges support from the Natural Sciences and Engineering Research Council of Canada (NSERC) grant RGPIN-2016-06718 and the Canada Research Chairs program. This research used the computational resources of the National Energy Research Scientific Computing Center, a DOE Office of Science User Facility supported by the Office of Science of the U.S. Department of Energy under Contract No. DE-AC02-05CH11231.

**REFERENCES**
1. Wright PE, Dyson HJ. Intrinsically disordered proteins in cellular signalling and regulation. *Nature Rev. Mol. Cell Bio.* **16**, 18-29 (2015).

2. Fawzi NL, Phillips AH, Ruscio JZ, Doucleff M, Wemmer DE, Head-Gordon T. Structure and dynamics of the Abeta(21-30) peptide from the interplay of NMR experiments and molecular simulations. *J. Am. Chem. Soc.* **130**, 6145-6158 (2008).

3. Gomes G-NW*, et al.* Conformational Ensembles of an Intrinsically Disordered Protein Consistent with NMR, SAXS, and Single-Molecule FRET. *J. Am. Chem. Soc.* **142**, 15697-15710 (2020).

4. Svergun D, Barberato C, Koch MHJ. CRYSOL - a Program to Evaluate X-ray Solution Scattering of Biological Macromolecules from Atomic Coordinates. *J. Appl. Cryst.* **28**, 768-773 (1995).

5. Bhowmick A*, et al.* Finding Our Way in the Dark Proteome. *J. Am. Chem. Soc.* **138**, 9730-9742 (2016).

6. Ozenne V*, et al.* Flexible-meccano: a tool for the generation of explicit ensemble descriptions of intrinsically disordered proteins and their associated experimental observables. *Bioinform.* **28**, 1463-1470 (2012).

7. Krzeminski M, Marsh JA, Neale C, Choy WY, Forman-Kay JD. Characterization of disordered proteins with ENSEMBLE. *Bioinform.* **29**, 398-399 (2013).

8. Ball KA, Wemmer DE, Head-Gordon T. Comparison of structure determination methods for intrinsically disordered amyloid-beta peptides. *J. Phys. Chem. B* **118**, 6405-6416 (2014).

9. Brookes DH, Head-Gordon T. Experimental Inferential Structure Determination of Ensembles for Intrinsically Disordered Proteins. *J. Am. Chem. Soc.* **138**, 4530-4538 (2016).

10. Kofinger J, Stelzl LS, Reuter K, Allande C, Reichel K, Hummer G. Efficient Ensemble Refinement by Reweighting. *J. Chem. Theo. Comput.* **15**, 3390-3401 (2019).

11. Lincoff J*, et al.* Extended experimental inferential structure determination method in determining the structural ensembles of disordered protein states. *Comm. Chem.* **3**, 74 (2020).

12. Nerenberg PS, Head-Gordon T. New developments in force fields for biomolecular simulations. *Curr. Opin. Struct. Biol.* **49**, 129-138 (2018).

13. Nerenberg PS, Head-Gordon T. Optimizing protein− solvent force fields to reproduce intrinsic conformational preferences of model peptides. *J. Chem. Theo. Comput.* **7**, 1220-1230 (2011).

14. Chong S-H, Chatterjee P, Ham S. Computer Simulations of Intrinsically Disordered Proteins. *Ann. Rev. Phys. Chem*. **68**, 117-134 (2017).

15. Henriques J, Cragnell C, Skepö M. Molecular dynamics simulations of intrinsically disordered proteins: force field evaluation and comparison with experiment. *J. Chem. Theo. Comput.* **11**, 3420-3431 (2015).

16. Siwy CM, Lockhart C, Klimov DK. Is the Conformational Ensemble of Alzheimer's Aβ10-40 Peptide Force Field Dependent? *PLOS Comput Bio* **13**, e1005314 (2017).

17. Wickstrom L, Okur A, Simmerling C. Evaluating the performance of the ff99SB force field based on NMR scalar coupling data. *Biophys. J.* **97**, 853-856 (2009).

18. Zaslavsky BY, Uversky VN. In Aqua Veritas: The Indispensable yet Mostly Ignored Role of Water in Phase Separation and Membrane-less Organelles. *Biochem.* **57**, 2437-2451 (2018).

19. Lindorff-Larsen K*, et al.* Improved side-chain torsion potentials for the Amber ff99SB protein force field. *Proteins* **78**, 1950-1958 (2010).

20. Robustelli P, Piana S, Shaw DE. Developing a molecular dynamics force field for both folded and disordered protein states. *Proc. Natl. Acad. Sci. USA* **115**, E4758-E4766 (2018).

21. Lincoff J, Sasmal S, Head-Gordon T. The combined force field-sampling problem in simulations of disordered amyloid-beta peptides. *J. Chem. Phys.* **150**, 104108 (2019).

22. Piana S, Donchev AG, Robustelli P, Shaw DE. Water dispersion interactions strongly influence simulated structural properties of disordered protein states. *J. Phys. Chem. B* **119**, 5113-5123 (2015).

23. Best RB, Zheng W, Mittal J. Balanced Protein–Water Interactions Improve Properties of Disordered Proteins and Non-Specific Protein Association. *J. Chem. Theo. Comput.* **10**, 5113-5124 (2014).

24. Abascal JLF, Vega C. A general purpose model for the condensed phases of water: TIP4P/2005. *J. Chem. Phys.* **123**, 234505 (2005).

25. Fluitt Aaron M, de Pablo Juan J. An Analysis of Biomolecular Force Fields for Simulations of Polyglutamine in Solution. *Biophys. J.* **109**, 1009-1018 (2015).

26. Wang Y, *et al.* Multiscaled exploration of coupled folding and binding of an intrinsically disordered molecular recognition element in measles virus nucleoprotein. *Proc. Natl. Acad. Sci. USA* **110**, E3743-E3752 (2013).

27. Moritsugu K, Terada T, Kidera A. Disorder-to-order transition of an intrinsically disordered region of sortase revealed by multiscale enhanced sampling. *J. Am. Chem. Soc.* **134**, 7094-7101 (2012).

28. Wells M, *et al.* Structure of tumor suppressor p53 and its intrinsically disordered N-terminal transactivation domain. *Proc. Natl. Acad. Sci. USA* **105**, 5762-5767 (2008).

29. Rauscher S, Pomès R. The liquid structure of elastin. *eLife* **6**, e26526 (2017).

30. Huang J, *et al.* CHARMM36m: an improved force field for folded and intrinsically disordered proteins. *Nature Meth.* **14**, 71-73 (2017).

31. Ponder JW, *et al.* Current status of the AMOEBA polarizable force field. *J. Phys. Chem. B* **114**, 2549 (2010).

32. Demerdash O, Wang L-P, Head-Gordon T. Advanced models for water simulations. *WIRES: Comput. Mol. Sci.* **8**, e1355 (2018).

33. Shi Y, *et al.* The Polarizable Atomic Multipole-based AMOEBA Force Field for Proteins. *J. Chem. Theo. Comput.* **9**, 4046-4063 (2013).

34. Ren P, Ponder JW. Polarizable atomic multipole water model for molecular mechanics simulation. *J. Phys. Chem. B* **107**, 5933-5947 (2003).

35. Maiorov VN, Crippen GM. Significance of Root-Mean-Square Deviation in Comparing Three-dimensional Structures of Globular Proteins. *J. Mol. Bio.* **235**, 625-634 (1994).

36. Zhou Y, Vitkup D, Karplus M. Native proteins are surface-molten solids: application of the lindemann criterion for the solid versus liquid state. *J. Mol. Bio.* **285**, 1371-1375 (1999).

37. Katava M, *et al.* Critical structural fluctuations of proteins upon thermal unfolding challenge the Lindemann criterion. *Proc. Natl. Acad. Sci. USA* **114**, 9361 (2017).

38. Cragnell C, Durand D, Cabane B, Skepö M. Coarse-grained modelling of the intrinsically disordered protein Histatin 5 in solution. Monte Carlo simulations in combination with SAXS. *Proteins: Struct, Func, Bioinform.* (2016).

39. Hickman AB, Klein DC, Dyda F. Melatonin Biosynthesis: The Structure of Serotonin N-Acetyltransferase at 2.5 A; Resolution Suggests a Catalytic Mechanism. *Mol. Cell* **3**, 23-32 (1999).

40. Tsunasawa S, Masaki T, Hirose M, Soejima M, Sakiyama F. The primary structure and structural characteristics of Achromobacter lyticus protease I, a lysine-specific serine protease. *J. Biol. Chem.* **264**, 3832-3839 (1989).

41. Dideberg O, *et al.* The crystal structure of the β-lactamase of Streptomyces albus G at 0.3 nm resolution. *Biochem. J.* **245**, 911-913 (1987).

42. Muller P, *et al.* The 1.70 angstroms X-ray crystal structure of Mycobacterium tuberculosis phosphoglycerate mutase. *Acta Cryst. D Biol. Cryst.* **61**, 309-315 (2005).

43. Paakkonen K, *et al.* Solution structures of the first and fourth TSR domains of F-spondin. *Proteins* **64**, 665-672 (2006).

44. Sutkeviciute I, *et al.* Unique DC-SIGN clustering activity of a small glycomimetic: A lesson for ligand design. *ACS Chem. Bio.* **9**, 1377-1385 (2014).

45. Ulrich A, Wahl MC. Structure and evolution of the spliceosomal peptidyl-prolyl cis-trans isomerase Cwc27. *Acta Cryst. D Biol. Cryst.* **70**, 3110-3123 (2014).

46. Wan Q, *et al.* Direct determination of protonation states and visualization of hydrogen bonding in a glycoside hydrolase with neutron crystallography. *Proc. Natl. Acad. Sci. USA* **112**, 12384-12389 (2015).

47. Kolinski A, Godzik A, Skolnick J. A general method for the prediction of the three dimensional structure and folding pathway of globular proteins: Application to designed helical proteins. *J. Chem. Phys.* **98**, 7420-7433 (1993).

48. Dima RI, Thirumalai D. Asymmetry in the Shapes of Folded and Denatured States of Proteins. *J. Phys. Chem. B* **108**, 6564-6570 (2004).

49. Yang L-W, Eyal E, Chennubhotla C, Jee J, Gronenborn AM, Bahar I. Insights into equilibrium dynamics of proteins from comparison of NMR and X-ray data with computational predictions. *Structure* **15**, 741-749 (2007).

50. Andrec M, Snyder DA, Zhou Z, Young J, Montelione GT, Levy RM. A large data set comparison of protein structures determined by crystallography and NMR: Statistical test for structural differences and the effect of crystal packing. *Proteins: Struct, Func, Bioinform.* **69**, 449-465 (2007).

51. Hura GL*, et al.* Small angle X-ray scattering-assisted protein structure prediction in CASP13 and emergence of solution structure differences. *Proteins: Struct, Func, Bioinform.* **87**, 1298-1314 (2019).

52. Pitera JW. Expected Distributions of Root-Mean-Square Positional Deviations in Proteins. *J. Phys. Chem. B* **118**, 6526-6530 (2014).

53. Lindemann F. The calculation of molecular vibration frequencies. *Z. Phys.* **11**, 609-612 (1910).

54. Zhou Y, Karplus M. Folding thermodynamics of a model three-helix-bundle protein. *Proc. Natl. Acad. Sci. USA* **94**, 14429-14432 (1997).

55. Lindorff-Larsen K, Piana S, Dror RO, Shaw DE. How Fast-Folding Proteins Fold. *Science* **334**, 517 (2011).

56. Alowolodu O, Johnson G, Alashwal L, Addou I, Zhdanova IV, Uversky VN. Intrinsic disorder in spondins and some of their interacting partners. *Intrinsically Disordered Proteins* **4**, e1255295-e1255295 (2016).

57. Brewer D, Hunter H, Lajoie G. NMR studies of the antimicrobial salivary peptides histatin 3 and histatin 5 in aqueous and nonaqueous solutions. *Biochem. Cell Bio.* **76**, 247-256 (1998).

58. Raj PA, Marcus E, Sukumaran DK. Structure of human salivary histatin 5 in aqueous and nonaqueous solutions. *Biopolymers* **45**, 51-67 (1998).

59. Shalongo W, Dugad L, Stellwagen E. Distribution of Helicity within the Model Peptide Acetyl(AAQAA)3amide. *J. Am. Chem. Soc.* **116**, 8288-8293 (1994).

60. Shirley WA, Brooks Iii CL. Curious structure in "canonical" alanine-based peptides. *Proteins: Struct, Func, Bioinform.* **28**, 59-71 (1997).

61. Huang J, MacKerell AD, Jr. Induction of peptide bond dipoles drives cooperative helix formation in the (AAQAA)3 peptide. *Biophys. J.* **107**, 991-997 (2014).

62. Mayo A, Yap K. Empirical analysis of backbone chemical shifts in proteins.( http://www.bmrb.wisc.edu/published/Ikura_cs_study) (2001).

63. Best RB, Hummer G. Optimized molecular dynamics force fields applied to the helix-coil transition of polypeptides. *J. Phys. Chem. B* **113**, 9004-9015 (2009).

64. Boonstra S, Onck PR, van der Giessen E. CHARMM TIP3P Water Model Suppresses Peptide Folding by Solvating the Unfolded State. *J. Phys. Chem. B* **120**, 3692-3698 (2016).

65. Marsh JA, Singh VK, Jia Z, Forman-Kay JD. Sensitivity of secondary structure propensities to sequence differences between α- and γ-synuclein: Implications for fibrillation. *Prot. Sci.* **15**, 2795-2804 (2006).

66. Camilloni C, De Simone A, Vranken WF, Vendruscolo M. Determination of Secondary Structure Populations in Disordered States of Proteins Using Nuclear Magnetic Resonance Chemical Shifts. *Biochem.* **51**, 2224-2231 (2012).

67. Hornak V, Abel R, Okur A, Strockbine B, Roitberg A, Simmerling C. Comparison of multiple Amber force fields and development of improved protein backbone parameters. *Proteins* **65**, 712-725 (2006).

68. Jorgensen WL, Chandrasekhar J, Madura JD, Impey RW, Klein ML. Comparison of simple potential functions for simulating liquid water. *J. Chem. Phys.* **79**, 926-935 (1983).

69. Horn HW, *et al.* Development of an improved four-site water model for biomolecular simulations: TIP4P-Ew. *J. Chem. Phys.* **120**, 9665-9678 (2004).

70. Eastman P, *et al.* OpenMM 4: a reusable, extensible, hardware independent library for high performance molecular simulation. *J. Chem. Theo. Comput.* **9**, 461 (2013).

71. Harger M, *et al.* Tinker-OpenMM: Absolute and relative alchemical free energies using AMOEBA on GPUs. *J. Comput. Chem.* **38**, 2047-2055 (2017).

72. Brown S, Head-Gordon T. Cool walking: A new Markov chain Monte Carlo sampling method. *J. Comput. Chem.* **24**, 68-76 (2003).

73. Lincoff J, Sasmal S, Head-Gordon T. Comparing generalized ensemble methods for sampling of systems with many degrees of freedom. *J. Chem. Phys.* **145**, 174107 (2016).

74. Roe DR, Cheatham TE, 3rd. PTRAJ and CPPTRAJ: Software for Processing and Analysis of Molecular Dynamics Trajectory Data. *J. Chem. Theo. Comput.* **9**, 3084-3095 (2013).

75. Jo S, Lim JB, Klauda JB, Im W. CHARMM-GUI Membrane Builder for mixed bilayers and its application to yeast membranes. *Biophys. J.* **97**, 50-58 (2009).

76. Rackers JA, *et al.* Tinker 8: Software Tools for Molecular Design. *J. Chem. Theo. Comput.* **14**, 5273-5289 (2018).